\documentclass[12pt,onecolumn,draftclsnofoot]{IEEEtran}
\renewcommand\IEEEkeywordsname{Index Terms}
\usepackage{blindtext}
\usepackage{graphicx}
\usepackage{amsmath}
\usepackage{amsfonts}
\usepackage{braket}
\usepackage{multirow}
\usepackage{subfigure}
\usepackage{cite}
\usepackage{xcolor}
\usepackage[ruled]{algorithm2e}
\usepackage[utf8]{inputenc}
\usepackage[english]{babel}
\usepackage{caption}
\usepackage{amsthm}
\usepackage{setspace}
\ifCLASSINFOpdf
\else
\fi
\newcommand{\vd}{{\bf d}}

\newcommand{\vg}{{\bf g}}
\newcommand{\vh}{{\bf h}}

\newcommand{\vu}{{\bf u}}

\newcommand{\vw}{{\bf w}}
\newcommand{\vx}{{\bf x}}
\newcommand{\vy}{{\bf y}}

\newcommand{\zv}{{\bf 0}}

\newcommand{\cT}{{\cal T}}

\newcommand{\bC}{{\mathbb{C}}}

\newcommand{\bN}{{\mathbb{N}}}

\newenvironment{mat}[1]{\left[\begin{array}{#1}}{\end{array}\right]}

\renewcommand{\tt}[2]{{\cT_{#1}^{#2}}}

\newcommand{\mA}{{\bf A}}

\newcommand{\mF}{{\bf F}}

\newcommand{\mH}{{\bf H}}

\newcommand{\mP}{{\bf P}}

\newcommand{\mW}{{\bf W}}

\newcommand{\ignore}[1]{}

\DeclareMathOperator{\diag}{diag}

\newcommand{\modd}[1]{\left\langle#1\right\rangle}

\begin{document}

%
\newtheorem{corol}[theorem]{Corollary}
\title{An Overview of Generalized Frequency Division Multiplexing (GFDM)}
%
%
%

\author{
Ching-Lun Tai$^1$, Tzu-Han Wang$^1$, and~Yu-Hua Huang$^1$\\
\quad\\
$^1$School of Electrical and Computer Engineering, Georgia Institute of Technology, GA, United States
}

\maketitle

\begin{abstract}
As a candidate waveform for next-generation wireless communications, generalized frequency division multiplexing (GFDM) features several decent properties which make it promising.
In this paper, we systematically overview the research about GFDM.
We start with GFDM transceivers with their main components, which consist of prototype filter design, low-complexity transceiver implementation, and symbol detection algorithms.
Then, we investigate a couple of non-ideal issues of GFDM, including synchronization issues, channel estimation, and in-phase/quadrature (I/Q) imbalance compensation.
Lastly, we study the applications of GFDM-based cognitive radio and full-duplex radio which boast of a high spectral efficiency.

\end{abstract}


\begin{IEEEkeywords}
Generalized frequency division multiplexing (GFDM), transceiver structure, non-ideal issue, cognitive radio, full-duplex radio
\end{IEEEkeywords}

%
\IEEEpeerreviewmaketitle

\section{Introduction}
\begin{figure}[h]
    \centering
    \includegraphics[width=15cm]{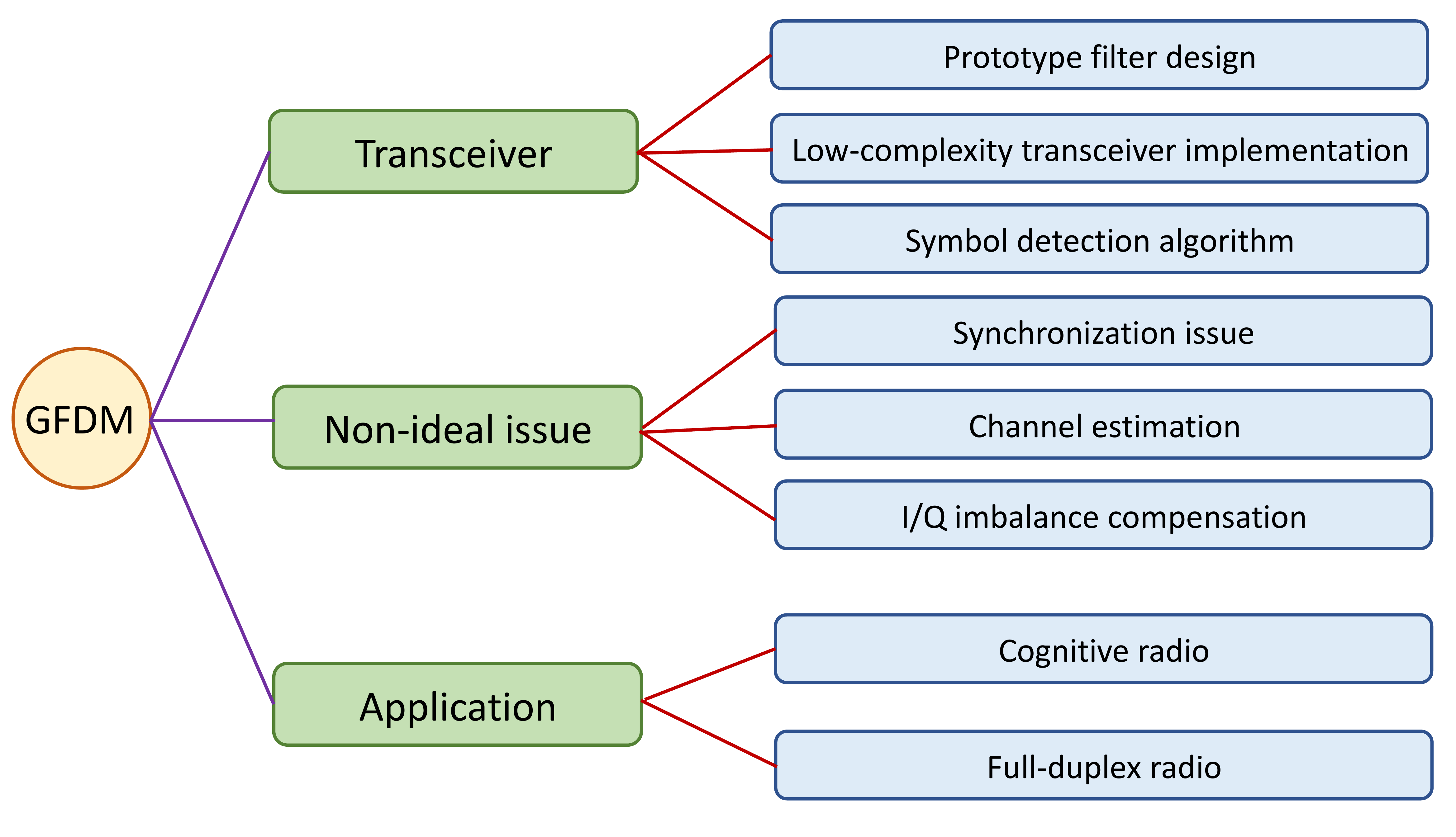}
    \caption{Organization of GFDM overview in this paper.}
    \label{fig:overview}
\end{figure}
Generalized frequency division multiplexing (GFDM) \cite{fettweis09} is a promising candidate waveform for next-generation wireless communications, featuring several advantages such as low latency, low peak-to-average ratio (PAPR), low out-of-band (OOB) emission and low adjacent channel leakage ratio (ACLR), and relaxed requirements of time and frequency synchronization \cite{michailow14}.

With a flexible transceiver structure, GFDM boasts of a high degree of freedom for transmitter and receiver design.
Therefore, GFDM is applicable to a variety of scenarios and its parameters can be adapted to meet the requirements of specific services.
In addition, there is abundant low-complexity implementation of GFDM transceivers, which are hence practical from an economic perspective, and various algorithms can be applied to the GFDM receiver, where the complexity is further reduced while maintaining an allowable performance.

In real-world applications, there are a couple of non-ideal issues that affect the performances of GFDM.
However, thanks to its flexible structure, plenty of techniques can be adopted for GFDM and help with the evaluation and compensation of the adverse effects caused by these issues.

Because of its decent properties, GFDM is suitable for a large number of applications.
Among all potential applications, cognitive radio and full-duplex radio are of great interest due to their high spectral efficiency, which help address the scarcity of bandwidth resources.
Accordingly, GFDM-based cognitive radio and full-duplex radio are considered an appealing solution to spectrum management in next-generation wireless communications.

In this paper, we provide a systematic overview of GFDM with the following topics (as summarized in Fig. \ref{fig:overview}):
\begin{itemize}
    \item \textbf{GFDM transceiver}: We briefly review the system model of GFDM. Besides, we study the important components of GFDM transceivers, including prototype filter design, low-complexity transceiver implementation, and symbol detection algorithms.
    \item \textbf{Non-ideal issues}: We investigate the effects and possible solutions of several non-ideal issues of GFDM, including synchronization issues, channel estimation, and in-phase/quadrature (I/Q) imbalance compensation.
    \item \textbf{GFDM-based cognitive radio and full-duplex radio}: We introduce the applications of cognitive radio and full-duplex radio and discuss about the research of GFDM-based cognitive radio and full-duplex radio.
\end{itemize}

The remainder of this paper is organized as follows.
Sec. \ref{sec:transceiver} introduces GFDM transceivers and their essential components.
The non-ideal issues of GFDM are investigated in Sec. \ref{sec:non-ideal}.
In Sec. \ref{sec:radio}, we study GFDM-based cognitive radio and full-duplex radio.
Finally, Sec. \ref{sec:conclusion} concludes the paper.

\textbf{\textit{Notations}}:
Boldfaced capital and lowercase letters denote matrices and column vectors, respectively. We use $\modd{.}_D$ to denote the modulo $D$.
We adopt the MATLAB subscripts $:$ and $a:b$ to denote all elements and the elements ordered from $a$ to $b$, respectively, of the subscripted objects.
Given a vector $\vu$, we use $[\vu]_n$ to denote the $n$th component of $\vu$ and $\diag(\vu)$ the diagonal matrix containing $\vu$ on its diagonal.  
Given a matrix $\mA$, we denote $[\mA]_{m, n}$, ${\mA}^T$, and ${\mA}^H$ its ($m$, $n$)th entry (zero-based indexing), transpose, and Hermitian transpose, respectively. 
We define $\zv_q$ to be the $q \times 1$ zero vector, and ${\mW}_q$ the normalized $q$-point discrete Fourier transform (DFT) matrix with ${[\mW_q]}_{m, n}=e^{-j2\pi mn/q}/\sqrt{q}, q \in \bN$.


\section{GFDM Transceiver}
\label{sec:transceiver}
\begin{figure}[h]
    \centering
    \includegraphics[width=15cm]{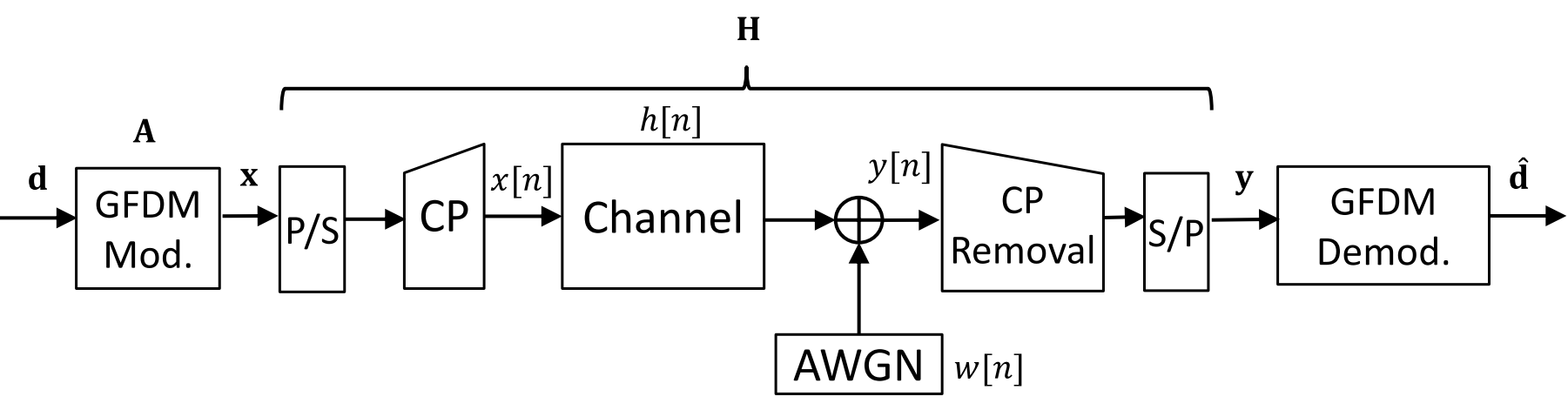}
    \caption{GFDM system model.}
    \label{fig:GFDM}
\end{figure}

GFDM is a block-based multicarrier communication scheme as shown in Fig. \ref{fig:GFDM} \cite{michailow14}.
Each GFDM block employs $K$ subcarriers, with each transmitting $M$ complex-valued subsymbols, and therefore a total of $D=KM$ symbols are transmitted within one block.

For a GFDM block $\vd\in\bC^D$, its $m$th subsymbol on the $k$th subcarrier $[\vd]_{k+mK}$ is pulse-shaped by a vector $\vg_{k,m}$, whose $n$th entry is $[\vg_{k,m}]_n=[\vg]_{\modd{n-mK}_D}e^{j2\pi kn/K},n=0,1,...,D-1,m=0,1,...,M-1,k=0,1,...,K-1$, where $\vg\in\bC^D$ is called the \emph{prototype filter} \cite{michailow14}.
Let
\begin{equation}
    \mA=[\vg_{0,0}...\vg_{K-1,0}\quad \vg_{0,1}...\vg_{K-1,1}\quad ...\quad \vg_{0,M-1}...\vg_{K-1,M-1}]
    \label{eq:A}
\end{equation}
be the GFDM transmitter matrix \cite{michailow14} and $\vx=\mA\vd$ be the transmit sample vector, whose $n$th entry is
\begin{equation}
    [\vx]_n=\sum_{k=0}^{K-1}\sum_{m=0}^{M-1}[\vd]_{k+mK}[\vg]_{\modd{n-mK}_D}e^{j2\pi kn/K}.
\end{equation}
Then, the vector $\vx$ passes through parallel-to-serial (P/S) conversion and is further added a cyclic prefix (CP) of length $L$, generating a GFDM digital baseband transmit signal $x[n]$.

Consider an $N$-tap wireless channel, which is a causal linear time-invariant  (LTI) system with impulse response $h[n]$, where $h[n]=0,n<0\ \mbox{or}\ n>N-1$.
Accordingly, the baseband receive signal is $y[n]=h[n]*x[n]+w[n]$, where $w[n]$ is the complex additive white Gaussian noise (AWGN) with variance $N_0$.
Denote $\vh=[h[0]\,h[1]...h[N-1]]^T$ and $\vw=[w[0]\,w[1]...w[D-1]]^T$.
Note that the combination of a CP and a linear convolution with a frequency-selective multipath channel can be modeled as a circular convolution \cite{tse05}.
Specifically, after CP removal and serial-to-parallel (S/P) conversion, the receive sample vector is obtained as
\begin{equation}
    \vy=\mH \vx+\vw=\mH \mA \vd+\vw=\mW_D^H\mbox{diag}(\mW_D\mA\vd)\mF_N\vh+\vw,
\end{equation}
where $\mF_N=[\sqrt{D}\mW_D]_{:,1:N}$, and $\mH\in\bC^{D \times D}$ is the circulant matrix whose first column is $[\vh^T \quad \zv_{D-N}^T]^T$.

Finally, the estimated GFDM block $\hat{\vd}$ is obtained after the demodulation process, which varies depending on the target applications.

\subsection{Prototype Filter Design}
For GFDM, the prototype filter $\vg$ determines the characteristics of the transmitter matrix $\mA$.
With the emergence of various services in a heterogeneous network, prototype filter design is critical in constructing a transmitter that meets the requirements of quality of service (QoS).
Classic prototype filters include the raised cosine (RC) filter featuring low inter subsymbol interference (ISI) \cite{michailow14} and Dirichlet filter (with its variants) causing no inter subcarrier interference (ICI) \cite{matthe14a,tai19}.
Over the past years, a variety of prototype filter design schemes have been proposed based on different considerations.

From an analytical perspective, the conditional number of a transmitter matrix is of great interest.
Specifically, the conditional number is related to the singularity \cite{lin16}, which affects the error rate, and noise enhancement factor (NEF) \cite{towliat20}, which indicates the level of noise enhancement, of a transmitter matrix.

When it comes to the radio resource allocation, the OOB radiation, which is related to ACLR and evaluates the energy leak experienced by the outer frequency band in terms of the power spectral density (PSD), becomes a main concern.

Considering the circuit implementation, one of the main focuses is the PAPR, which is defined as $\mbox{PAPR}(\vx)=(\underset{n}{\mbox{max}}|[\vx]_n|^2)/(\sum_{n=0}^{D-1}|[\vx]_n|^2/D)\geq 1$ for a transmit sample vector $\vx$.
A larger PAPR implies that the transmit samples are more subject to the distortion caused by a power amplifier (PA).

Another practical concern is the transmission rate, which determines the number of symbols transmitted within a fixed time period.

In the existing literature, \cite{yoshizawa16} and \cite{nimr17} propose their design based on the conditional number, adopting a frequency shift technique and radix-2 fast Fourier transform (FFT), respectively.
With a different focus, \cite{sim19} and \cite{liu19} design their prototype filters by solving optimization problems which minimize the PAPR.
Instead of dealing with only one consideration, some works propose their prototype filter design with multiple considerations taken into account.
\cite{han17} proposes an algorithm to jointly optimize the OOB radiation and transmission rate; \cite{chen17a} solves an optimization problem that minimizes the OOB radiation while satisfying an NEF constraint; \cite{tai18} minimizes the approximate PAPR while considering the NEF and OOB radiation by solving an optimization problem.  



\subsection{Low-complexity Transceiver Implementation}
In order to achieve low-cost realizations of GFDM for real-world applications, the complexity, which is defined as the number of complex multiplications (CMs) involved in both modulation at the transmitter and demodulation at the receiver, of GFDM transceivers is of great importance.
Note that the complexity is closely related to the amount of computing resources consumed, the data processing time an application takes, and the latency a user experiences.
With the goal of fast processing and low latency in next-generation wireless communications \cite{popovski14}, low-complexity transceiver implementation is a critical topic for GFDM researchers.

For transceiver design, it could focus on a single side (i.e., the transmitter or receiver side), or it could jointly consider both transmitter and receiver sides.

The transmitter is mostly characterized by the transmitter matrix $\mA$.
Although the matrix $\mA$ is fully determined once the prototype filter $\vg$ is explicitly specified, the complexity involved in obtaining the transmit sample vector $\vx$, which is equal to $\mA\vd$, varies depending on the computing process.
The main reason is that the matrix $\mA$ can be decomposed into various forms by using techniques which transform a matrix into a cascade of smaller sparse matrices and therefore reduce the complexity.

Unlike the transmitter, the receiver structure is more flexible and is constructed according to the specifications of target applications.
Among all types of receivers, linear receivers are the ones that are commonly used for the low-complexity purpose, including the famous zero forcing (ZF) receiver and linear minimum mean square error (LMMSE) receiver \cite{michailow14,matthe14b}.
ZF receivers reverse the cascaded operations performed by the transmitter matrix $\mA$ and the channel circulant matrix $\mH$, featuring a lower complexity but a poorer performance.
On the other hand, LMMSE receivers require the statistics of noises to minimize the mean square error between the original and estimated GFDM blocks, $\vd$ and $\hat{\vd}$, resulting in a higher complexity but a better performance.

Based on the above guidelines and other advanced techniques, there are plenty of low-complexity transceiver schemes proposed for a more efficient implementation.
Several works focus on the receiver side.
\cite{gaspar13} and \cite{dias19} reduce the receiver complexity by employing frequency domain processing.
In addition, different techniques have been adopted to achieve low-complexity receiver implementation, such as sparsification and block diagonalization \cite{farhang15}, Taylor series and conjugate gradient \cite{tiwari18}, and tabu search \cite{jeong19}.
Differently, some works focus on the transmitter side, such as \cite{lin15} reducing the transmitter complexity with the technique of subcarrier-wise DFT.
Combining both transmitter and receiver sides, several low-complexity schemes have been proposed over the whole transceiver.
\cite{farhang16} reduces the complexity of both the transmitter and receiver by taking advantage of the sparsity in modulation and the block circulant property in demodulation.
Unitary transmitter matrices are adopted in \cite{chen17b} for low-complexity transceiver implementation.
In \cite{nimr19}, only partial subcarriers are allocated in order to reduce the complexity of the whole transceiver.

\subsection{Symbol Detection Algorithms}
Apart from conventional demodulation processes (e.g., linear receivers), there are a variety of algorithms which can be adopted to provide the estimated GFDM block $\hat{\vd}$, and these algorithms are often called symbol detectors.

One of the largest differences between conventional demodulation processes and symbol detection algorithms is the tradeoff between complexity and performances.
Conventional demodulation processes suffer from a higher complexity but provide a (statistically) better performance, while symbol detection algorithms feature a lower complexity but offer an inferior performance.
Therefore, symbol detection algorithms are more suitable for applications where a large number of subcarriers $K$ and/or a large number of subsymbols (within a subcarrier) $M$ are used and the complexity of conventional demodulation processes is prohibitively high.

Particularly, symbol detection algorithms work for multiple-input-multiple-output (MIMO) systems, where multiple transmitters and multiple receivers are used to provide a benefit of both diversity gains (due to the multiple paths created in the system) and multiplexing gains (due to the spatial correlations between transmitters or between receivers) \cite{zheng03}.
Note that the combination of MIMO and GFDM (i.e., MIMO-GFDM) often leads to a large block size where conventional demodulation processes become infeasible and only symbol detection algorithms work (despite the sub-optimal performance).

Several symbol detection algorithms have been adopted for GFDM in the existing literature.
For instance, \cite{zhang17b} and \cite{turhan19} use orthogonal approximate message passing (OAMP) and deep convolutional neural networks (CNNs) to achieve GFDM symbol detection.
A couple of symbol detection algorithms have been employed for MIMO-GFDM systems, such as the Markov chain Monte Carlo (MCMC) algorithm \cite{zhang15}, MMSE sorted QR-decomposition (SQRD) with sphere decoding (SD) \cite{matthe15b}, expectation propagation (EP) \cite{zhang16}, and MMSE parallel interference cancellation (PIC) \cite{matthe18}.

\section{Non-Ideal Issues of GFDM}
\label{sec:non-ideal}
Within a synthetic simulated environment, we can evaluate the performance limit of a communication system over an ideal case.
However, in real-world applications, there are a variety of non-ideal issues that deteriorate the performances of a communication system.
Accordingly, the performances of a communication system in practical use are expected to be worse than those observed in the ideal case, which provides an upper bound of performances.

In the following, we investigate the effects and possible solutions of three typical non-ideal issues, including synchronization issues, channel estimation, and I/Q imbalance compensation, of GFDM.

\subsection{Synchronization Issues}
\begin{figure}[h]
    \centering
    \includegraphics[width=15cm]{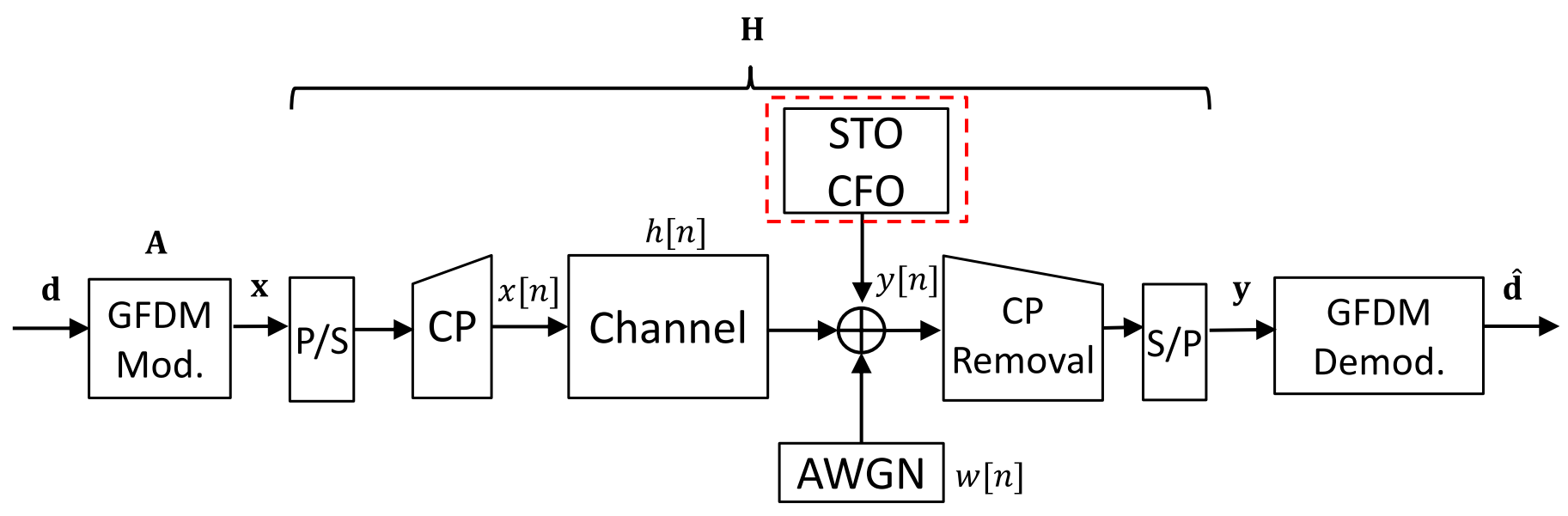}
    \caption{An illustration of occurrence of synchronization errors (red dashed box indicates the additional block compared with Fig. \ref{fig:GFDM}).}
    \label{fig:sync}
\end{figure}
In the ideal case, a perfect synchronization is assumed at the GFDM receiver, where the demodulation process works.
However, in practical use, there are plenty of factors that cause a synchronization error, which leads to a subcarrier/subsymbol misalignment at the receiver and thus significantly increases the error rate \cite{gaspar17,sharma19}.
An illustration of the occurrence of synchronization errors is as shown in Fig. \ref{fig:sync}.

Basically, there are two types of synchronization errors, including symbol time offset (STO) and carrier frequency offset (CFO), which correspond to the asynchronization in time domain and frequency domain, respectively \cite{michailow14}.
The main reasons of their occurrence include the Doppler effect caused by the mobility of users and the multipath effect caused by the obstacles in the environment.
In order to compensate the adverse effects of synchronization errors, it is critical to estimate them.
Typical approaches include a supervised estimation with the use of training sequences and a unsupervised (blind) estimation with the use of statistical methods.

A more complicated scenario happens in an uplink multiuser system, where multiple users transmit their data to the base station. 
In addition to synchronization errors, there is also multiuser interference (MUI) that further deteriorates the demodulation performances.
For this case, MUI cancellation methods, which eliminate synchronization errors simultaneously, are essential.

Accordingly, a couple of schemes working on synchronization issues have been proposed.
For a supervised estimation of synchronization errors, different structures for training sequences have been used, such as embedded midamble \cite{gaspar15c}, scattered pilots \cite{matthe15a}, pseudo-circular preamble \cite{gaspar15b}, and partial employed subcarriers \cite{lee16}, where training symbols are put in the middle, a specific position, the front, or all positions of a subcarrier, respectively.
In addition, \cite{wang16a} proposes an unsupervised (blind) estimation scheme which derives from the statistical maximum-likelihood (ML) method. 
For an uplink multiuser system, synchronization errors and MUI need to be jointly addressed.
While \cite{lim18} takes an approach of maximizing signal-to-interference ratio (SIR), \cite{lim19} adopts the techniques of weighted parallel interference cancellation (WPIC) and adaptive interference cancellation filter (AICF).

\subsection{Channel Estimation}
\begin{figure}[h]
    \centering
    \includegraphics[width=15cm]{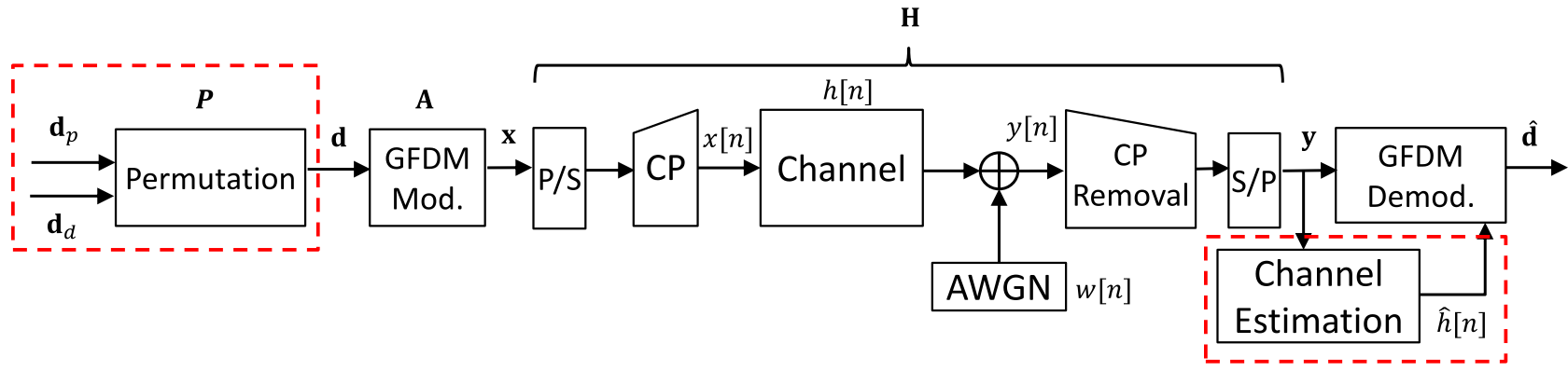}
    \caption{An illustration of channel estimation (red dashed boxes indicate the additional blocks compared with Fig. \ref{fig:GFDM}).}
    \label{fig:chan_esti}
\end{figure}
Within an ideal environment, the full channel state information (CSI) about the channel vector $\vh$ is assumed to be known at the GFDM receiver (which is called Genie-aided condition).
However, actually the CSI is hardly known in advance, i.e., such information is seldom a prior knowledge, in real-world applications.

Since the receiver is built upon the knowledge of CSI, the channel estimation is an essential step before the demodulation process takes place.
An illustration of channel estimation is as shown in Fig. \ref{fig:chan_esti}.
In order to achieve channel estimation, we require the pilots, which are symbols with fixed values that are generated for the estimation purpose, be contained in the GFDM block $\vd$.
Specifically, we create a concatenation of two subvectors, including the pilot vector $\vd_p$ which represents the pilots for channel estimation and the data vector $\vd_d$ which represents the data symbols to be transmitted, and permute it with a permutation matrix $\mP$ (which is equivalent to a linear combination of $\vd_p$ and $\vd_d$), generating the resulting GFDM block $\vd$, i.e., $\vd=\mP[\vd_p^T\ \vd_d^T]^T$.

As a non-orthogonal waveform, GFDM suffers from potential ICI (which can be avoided if ICI-free prototype filters are used, e.g., the Dirichlet filter) and inherent ISI, resulting in a unique challenge for channel estimation.
In order to deal with the effects of interference, the pilot structure and channel estimator design become the main focuses of GFDM channel estimation.
There exist various pilot structures that can be adopted for GFDM channel estimation, including scattered pilots where pilots are evenly scattered throughout the block, preamble/postamble pilots where pilots are in the front/back of the subcarrier, and pilot subcarriers where all positions of selected subcarriers are reserved for pilots.
Besides, there are a variety of channel estimators that one can choose from, including the famous linear channel estimators, e.g., the least square (LS) estimator and LMMSE estimator.

Based on different pilot structures and channel estimators, a couple of schemes have been proposed to achieve channel estimation.
Several schemes work on scattered pilots, including \cite{vilaipornsawai14} adopting the matched filter as the channel estimator, and \cite{akai17,ehsanfar19,tai20} performing channel estimation with the use of the LMMSE estimator.
In addition, \cite{ehsanfar16b} and \cite{ehsanfar17} both employ linear estimators (including the LS and LMMSE estimators), but selecting preamble/postamble pilots and pilot subcarriers, respectively, as the pilot structures.

\subsection{I/Q Imbalance Compensation}
\begin{figure}[h]
    \centering
    \includegraphics[width=15cm]{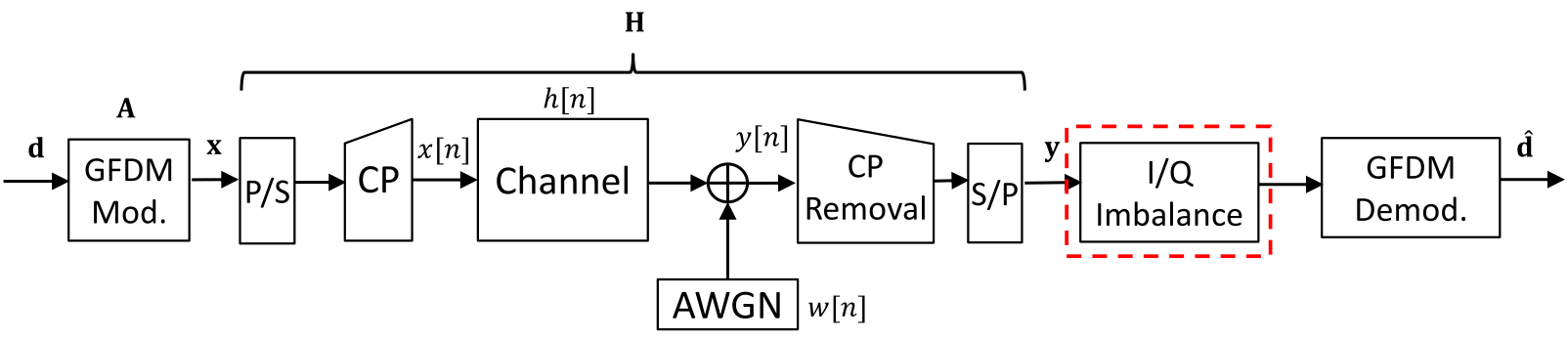}
    \caption{An illustration of occurrence of I/Q imbalance (red dashed box indicates the additional block compared with Fig. \ref{fig:GFDM}).}
    \label{fig:I/Q}
\end{figure}
Usually under an ideal setting, it is assumed that there is no RF impairment in GFDM circuit implementation.
Nonetheless, RF impairments, such as phase noises occurring at the receiver, are inevitable issues when it comes to real-world applications.

From an economic perspective toward the design of next-generation wireless communications, direct-conversion transceivers are an appealing option due to its decent properties such as small size, low cost, and low energy consumption (low power) \cite{mirabbasi00,pan12}, involving only a single mixing stage.
However, there are several RF impairment issues regarding direct-conversion transceivers, and one of them is the I/Q imbalance, which indicates the misalignment (in amplitudes and/or phases) between the in-phase and quadrature paths over the circuit implementation.
An illustration of the occurrence of I/Q imbalance is as shown in Fig. \ref{fig:I/Q}.

In order to address I/Q imbalance, it is critical to estimate its level and design the corresponding I/Q imbalance compensation schemes.
A couple of works have proposed their schemes about analyzing and compensating the effects of I/Q imbalance.
In \cite{lupupa15}, the authors analyze the I/Q imbalance of GFDM under Weibull fading, which is a statistical model for wireless indoor/outdoor channels.
In order to compensate I/Q imbalance, \cite{tang17b} and \cite{cheng18} propose a supervised scheme with the use of training sequences and a unsupervised (blind) scheme adopting statistical methods, respectively, to estimate the level of I/Q imbalance.

\section{GFDM-Based Cognitive Radio and Full-Duplex Radio}
\label{sec:radio}
Due to its decent properties such as low OOB radiation and low PAPR, GFDM is a promising solution to many applications required in next-generation wireless communications.
Among all potential applications, we investigate two of them, including the cognitive radio and the full-duplex radio, which serve as a remedy for the scarcity of bandwidth resources and the congestion in the frequency band.

Cognitive radio is a technique of dynamic bandwidth resource management, where unlicensed opportunistic users (also known as secondary users) detect unused channels (which are not occupied by their licensed primary users) and access them, resulting in a higher spectral efficiency.
The critical challenges of the cognitive radio include the OOB radiation and ICI, which are caused by secondary users and affect the primary users in adjacent channels.
In this context, GFDM is suitable for cognitive radio since it features low OOB radiation and its ICI can be eliminated with interference cancellation techniques or with the use of ICI-free prototype filters (e.g., the Dirichlet filter).
Recent research about GFDM-based cognitive radio mainly focuses on further reduction of OOB radiation and interference cancellation techniques.

Full-duplex radio is a duplex technique which allows the simultaneous mutual transmission and reception between two devices in a point-to-point communication, featuring a doubled system capacity and a higher spectral efficiency due to the feasibility of bidirectional communication.
The key challenge of full-duplex radio is the self-interference, where the transmission and reception processes interfere the operations of each other due to the signal leakage resulting from the deficiency of RF circuits.
Although GFDM suffers from inherent ISI, its nice characteristics such as low OOB radiation and low PAPR still make itself an appealing choice for full-duplex radio, since these two characteristics mitigate the design complexity of RF circuits. 
The main focus of recent research about GFDM-based full-duplex radio is the self-interference cancellation techniques.

Combining both cognitive radio and full-duplex radio, the full-duplex cognitive radio integrates the advantages and limits of two radio schemes, featuring an ultra-high spectral efficiency but facing challenges such as the OOB radiation and self/adjacent interference.
Full-duplex cognitive radio is a rather new concept for GFDM, and the research about GFDM full-duplex cognitive radio is around the problems derived from the combination of two radio schemes.

In the existing literature, there are several works dedicated to the realizations of GFDM-based cognitive radio and full-duplex radio.
Focusing on GFDM-based cognitive radio, \cite{datta12a} and \cite{datta14b} cancel the ICI with a double-sided serial interference cancellation technique and by inserting additional subcarriers, respectively.
Moreover, several related issues such as the detection of unused spectrum \cite{panaitopol12}, experimental testbeds \cite{danneberg14a}, and power allocation along with PA non-linearity alleviation \cite{mohammadian19} have been addressed over the past years.
For GFDM-based full-duplex radio, both \cite{mohammadian19b} and \cite{mohammadian20} deal with the self-interference cancellation when non-ideal issues such as synchronization errors and I/Q imbalance are present.
Extended from \cite{mohammadian19b} and \cite{mohammadian20}, the authors of \cite{mohammadian19c} focus on the self/adjacent interference cancellation of GFDM-based full-duplex cognitive radio in the presence of non-ideal issues.

\section{Conclusion}
\label{sec:conclusion}
In this paper, we present a systematic overview of GFDM, covering three main topics: GFDM transceivers, non-ideal issues of GFDM, and GFDM-based cognitive radio and full-duplex radio.
For GFDM transceivers, we provide a brief review on the GFDM system model and introduce their essential components, which consist of prototype filter design, low-complexity transceiver implementation, and symbol detection algorithms.
In addition, we investigate non-ideal issues of GFDM, including synchronization issues, channel estimation, and I/Q imbalance compensation.
Lastly, we study the applications of GFDM-based cognitive radio and full-duplex radio, pointing out the research challenges and recent advancements involved.


%




\ifCLASSOPTIONcaptionsoff
  \newpage
\fi



%


\bibliographystyle{IEEEtran}
\bibliography{IEEEabrv,waveform}

\begin{thebibliography}{10}
\providecommand{\url}[1]{#1}
\csname url@samestyle\endcsname
\providecommand{\newblock}{\relax}
\providecommand{\bibinfo}[2]{#2}
\providecommand{\BIBentrySTDinterwordspacing}{\spaceskip=0pt\relax}
\providecommand{\BIBentryALTinterwordstretchfactor}{4}
\providecommand{\BIBentryALTinterwordspacing}{\spaceskip=\fontdimen2\font plus
\BIBentryALTinterwordstretchfactor\fontdimen3\font minus
  \fontdimen4\font\relax}
\providecommand{\BIBforeignlanguage}[2]{{%
\expandafter\ifx\csname l@#1\endcsname\relax
\typeout{** WARNING: IEEEtran.bst: No hyphenation pattern has been}%
\typeout{** loaded for the language `#1'. Using the pattern for}%
\typeout{** the default language instead.}%
\else
\language=\csname l@#1\endcsname
\fi
#2}}
\providecommand{\BIBdecl}{\relax}
\BIBdecl

\bibitem{fettweis09}
G.~Fettweis, M.~Krondorf, and S.~Bittner, ``{GFDM - Generalized Frequency
  Division Multiplexing},'' in \emph{Veh. Technol. Conf., 2009. VTC Spring
  2009. IEEE 69th}, Apr. 2009, pp. 1--4.

\bibitem{michailow14}
N.~{Michailow}, M.~{Matthé}, I.~S. {Gaspar}, A.~N. {Caldevilla}, L.~L.
  {Mendes}, A.~{Festag}, and G.~{Fettweis}, ``{Generalized Frequency Division
  Multiplexing for 5th Generation Cellular Networks},'' \emph{IEEE Transactions
  on Communications}, vol.~62, no.~9, pp. 3045--3061, 2014.

\bibitem{tse05}
D.~Tse and P.~Viswanath, \emph{{Fundamentals of wireless communication}}.\hskip
  1em plus 0.5em minus 0.4em\relax Cambridge university press, 2005.

\bibitem{matthe14a}
M.~Matth\'{e}, N.~Michailow, I.~Gaspar, and G.~Fettweis, ``{Influence of pulse
  shaping on bit error rate performance and out of band radiation of
  Generalized Frequency Division Multiplexing},'' in \emph{Proc. IEEE ICC
  Workshop}, 2014, pp. 43--48.

\bibitem{tai19}
C.~{Tai}, B.~{Su}, and C.~{Jia}, ``{Frequency-domain Decoupling for MIMO-GFDM
  Spatial Multiplexing},'' in \emph{ICASSP 2019 - 2019 IEEE International
  Conference on Acoustics, Speech and Signal Processing (ICASSP)}, 2019, pp.
  4799--4803.

\bibitem{lin16}
D.~W. Lin and P.~S. Wang, ``On the configuration-dependent singularity of
  {GFDM} pulse-shaping filter banks,'' \emph{{IEEE} Commun. Lett.}, vol.~20,
  no.~10, pp. 1975--1978, Oct. 2016.

\bibitem{towliat20}
M.~{Towliat}, M.~{Rajabzadeh}, and S.~M. J.~A. {Tabatabaee}, ``{On the Noise
  Enhancement of GFDM},'' \emph{IEEE Wireless Communications Letters}, pp.
  1--1, 2020.

\bibitem{yoshizawa16}
A.~{Yoshizawa}, R.~{Kimura}, and R.~{Sawai}, ``{A Singularity-Free GFDM
  Modulation Scheme with Parametric Shaping Filter Sampling},'' in \emph{2016
  IEEE 84th Vehicular Technology Conference (VTC-Fall)}, 2016, pp. 1--5.

\bibitem{nimr17}
A.~{Nimr}, M.~{Matthé}, D.~{Zhang}, and G.~{Fettweis}, ``{Optimal Radix-2 FFT
  Compatible Filters for GFDM},'' \emph{IEEE Communications Letters}, vol.~21,
  no.~7, pp. 1497--1500, 2017.

\bibitem{sim19}
Z.~A. {Sim}, R.~{Reine}, Z.~{Zang}, F.~H. {Juwono}, and L.~{Gopal}, ``{Reducing
  the PAPR of GFDM Systems with Quadratic Programming Filter Design},'' in
  \emph{2019 IEEE 89th Vehicular Technology Conference (VTC2019-Spring)}, 2019,
  pp. 1--5.

\bibitem{liu19}
K.~{Liu}, W.~{Deng}, and Y.~{Liu}, ``{Theoretical Analysis of the
  Peak-to-Average Power Ratio and Optimal Pulse Shaping Filter Design for GFDM
  Systems},'' \emph{IEEE Transactions on Signal Processing}, vol.~67, no.~13,
  pp. 3455--3470, 2019.

\bibitem{han17}
S.~{Han}, Y.~{Sung}, and Y.~H. {Lee}, ``{Filter Design for Generalized
  Frequency-Division Multiplexing},'' \emph{IEEE Transactions on Signal
  Processing}, vol.~65, no.~7, pp. 1644--1659, 2017.

\bibitem{chen17a}
P.~{Chen} and B.~{Su}, ``{Filter optimization of out-of-band radiation with
  performance constraints for GFDM systems},'' in \emph{2017 IEEE 18th
  International Workshop on Signal Processing Advances in Wireless
  Communications (SPAWC)}, 2017, pp. 1--5.

\bibitem{tai18}
C.~L. Tai, B.~Su, and P.~C. Chen, ``Optimal filter design for gfdm that
  minimizes papr under performance constraints,'' in \emph{2018 IEEE Wireless
  Communications and Networking Conference (WCNC)}, April 2018, pp. 1--6.

\bibitem{popovski14}
P.~{Popovski}, ``{Ultra-reliable communication in 5G wireless systems},'' in
  \emph{1st International Conference on 5G for Ubiquitous Connectivity}, 2014,
  pp. 146--151.

\bibitem{matthe14b}
M.~Matthé, L.~Mendes, and G.~Fettweis, ``{Generalized Frequency Division
  Multiplexing in a {G}abor Transform Setting},'' \emph{{IEEE} Commun. Lett.},
  vol.~18, no.~8, pp. 1379--1382, Aug. 2014.

\bibitem{gaspar13}
I.~Gaspar, N.~Michailow, A.~Navarro, E.~Ohlmer, S.~Krone, and G.~Fettweis,
  ``Low complexity gfdm receiver based on sparse frequency domain processing,''
  in \emph{Veh. Technol. Conf. (VTC Spring), 2013 IEEE 77th}, June 2013, pp.
  1--6.

\bibitem{dias19}
W.~D. {Dias}, L.~L. {Mendes}, and J.~J. P.~C. {Rodrigues}, ``{Low Complexity
  GFDM Receiver for Frequency-Selective Channels},'' \emph{IEEE Communications
  Letters}, vol.~23, no.~7, pp. 1166--1169, 2019.

\bibitem{farhang15}
A.~Farhang, N.~Marchetti, and L.~E. Doyle, ``Low complexity gfdm receiver
  design: A new approach,'' in \emph{2015 IEEE Int. Conf. on Commun. (ICC)},
  June 2015, pp. 4775--4780.

\bibitem{tiwari18}
S.~{Tiwari} and S.~S. {Das}, ``{Low-Complexity Joint-MMSE GFDM Receiver},''
  \emph{IEEE Transactions on Communications}, vol.~66, no.~4, pp. 1661--1674,
  2018.

\bibitem{jeong19}
J.~{Jeong}, I.~{Jung}, J.~{Kim}, and D.~{Hong}, ``{A New GFDM Receiver with
  Tabu Search},'' in \emph{2019 IEEE 89th Vehicular Technology Conference
  (VTC2019-Spring)}, 2019, pp. 1--5.

\bibitem{lin15}
H.~Lin and P.~Siohan, ``Orthogonality improved {GFDM} with low complexity
  implementation,'' in \emph{2015 IEEE Wireless Commun. and Networking Conf.
  (WCNC)}, Mar. 2015, pp. 597--602.

\bibitem{farhang16}
A.~Farhang, N.~Marchetti, and L.~E. Doyle, ``Low-complexity modem design for
  {GFDM},'' \emph{{IEEE} Trans. Signal Process.}, vol.~64, no.~6, pp.
  1507--1518, Mar. 2016.

\bibitem{chen17b}
P.~C. Chen, B.~Su, and Y.~Huang, ``{Matrix Characterization for GFDM: Low
  Complexity MMSE Receivers and Optimal Filters},'' \emph{IEEE Transactions on
  Signal Processing}, vol.~65, no.~18, pp. 4940--4955, Sept 2017.

\bibitem{nimr19}
A.~{Nimr}, M.~{Chafii}, and G.~{Fettweis}, ``{Low-Complexity Transceiver for
  GFDM systems with Partially Allocated Subcarriers},'' in \emph{2019 IEEE
  Wireless Communications and Networking Conference (WCNC)}, 2019, pp. 1--6.

\bibitem{zheng03}
{Lizhong Zheng} and D.~N.~C. {Tse}, ``{Diversity and multiplexing: a
  fundamental tradeoff in multiple-antenna channels},'' \emph{IEEE Transactions
  on Information Theory}, vol.~49, no.~5, pp. 1073--1096, 2003.

\bibitem{zhang17b}
S.~{Zhang}, C.~{Wen}, K.~{Takeuchi}, and S.~{Jin}, ``{Orthogonal approximate
  message passing for GFDM detection},'' in \emph{2017 IEEE 18th International
  Workshop on Signal Processing Advances in Wireless Communications (SPAWC)},
  2017, pp. 1--5.

\bibitem{turhan19}
M.~{Turhan}, E.~{Öztürk}, and H.~A. {Çırpan}, ``{Deep Convolutional
  Learning-Aided Detector for Generalized Frequency Division Multiplexing with
  Index Modulation},'' in \emph{2019 IEEE 30th Annual International Symposium
  on Personal, Indoor and Mobile Radio Communications (PIMRC)}, 2019, pp. 1--6.

\bibitem{zhang15}
D.~Zhang, M.~Matthé, L.~L. Mendes, and G.~Fettweis, ``{A Markov chain Monte
  Carlo algorithm for near-optimum detection of MIMO-GFDM signals},'' in
  \emph{Personal, Indoor, and Mobile Radio Commun. (PIMRC), 2015 IEEE 26th
  Annual Int. Symposium on}, Aug 2015, pp. 281--286.

\bibitem{matthe15b}
M.~Matthé, I.~Gaspar, D.~Zhang, and G.~Fettweis, ``{Near-{ML} Detection for
  {MIMO}-{GFDM}},'' in \emph{Veh. Technol. Conf. (VTC Fall), 2015 IEEE 82nd},
  Sep. 2015, pp. 1--2.

\bibitem{zhang16}
D.~Zhang, L.~L. Mendes, M.~Matthé, I.~S. Gaspar, N.~Michailow, and G.~P.
  Fettweis, ``{Expectation Propagation for Near-Optimum Detection of MIMO-GFDM
  Signals},'' \emph{{IEEE} Trans. Wireless Commun.}, vol.~15, no.~2, pp.
  1045--1062, Feb 2016.

\bibitem{matthe18}
M.~{Matthé}, D.~{Zhang}, and G.~{Fettweis}, ``{Low-Complexity Iterative
  MMSE-PIC Detection for MIMO-GFDM},'' \emph{IEEE Transactions on
  Communications}, vol.~66, no.~4, pp. 1467--1480, 2018.

\bibitem{gaspar17}
D.~{Gaspar}, L.~{Mendes}, and T.~{Pimenta}, ``{GFDM BER Under Synchronization
  Errors},'' \emph{IEEE Communications Letters}, vol.~21, no.~8, pp.
  1743--1746, 2017.

\bibitem{sharma19}
N.~{Sharma}, A.~{Kumar}, M.~{Magarini}, S.~{Bregni}, and D.~N.~K. {Jayakody},
  ``{Impact of CFO on Low Latency-Enabled UAV Using "Better Than Nyquist" Pulse
  Shaping in GFDM},'' in \emph{2019 IEEE 89th Vehicular Technology Conference
  (VTC2019-Spring)}, 2019, pp. 1--6.

\bibitem{gaspar15c}
I.~Gaspar and G.~Fettweis, ``An embedded midamble synchronization approach for
  generalized frequency division multiplexing,'' in \emph{2015 IEEE Global
  Commun. Conf. (GLOBECOM)}, Dec. 2015, pp. 1--5.

\bibitem{matthe15a}
M.~Matthé, L.~L. Mendes, and G.~Fettweis, ``Asynchronous multi-user uplink
  transmission with generalized frequency division multiplexing,'' in
  \emph{2015 IEEE Int. Conf. on Commun. Workshop (ICCW)}, Jun. 2015, pp.
  2269--2275.

\bibitem{gaspar15b}
I.~Gaspar, A.~Festag, and G.~Fettweis, ``Synchronization using a
  pseudo-circular preamble for generalized frequency division multiplexing in
  vehicular communication,'' in \emph{Veh. Technol. Conf. (VTC Fall), 2015 IEEE
  82nd}, Sep. 2015, pp. 1--5.

\bibitem{lee16}
K.~{Lee}, M.~{Kang}, E.~{Jeong}, D.~{Park}, and Y.~H. {Lee}, ``{Use of training
  subcarriers for synchronization in low latency uplink communication with
  GFDM},'' in \emph{2016 IEEE 17th International Workshop on Signal Processing
  Advances in Wireless Communications (SPAWC)}, 2016, pp. 1--6.

\bibitem{wang16a}
P.~S. Wang and D.~W. Lin, ``Maximum-likelihood blind synchronization for {GFDM}
  systems,'' \emph{{IEEE} Signal Process. Lett.}, vol.~23, no.~6, pp. 790--794,
  Jun. 2016.

\bibitem{lim18}
B.~{Lim} and Y.~{Ko}, ``{Optimal receiver filter for GFDM with timing and
  frequency offsets in uplink multiuser systems},'' in \emph{2018 IEEE Wireless
  Communications and Networking Conference (WCNC)}, 2018, pp. 1--6.

\bibitem{lim19}
------, ``{Multiuser Interference Cancellation for GFDM With Timing and
  Frequency Offsets},'' \emph{IEEE Transactions on Communications}, vol.~67,
  no.~6, pp. 4337--4349, 2019.

\bibitem{vilaipornsawai14}
U.~Vilaipornsawai and M.~Jia, ``{Scattered-pilot channel estimation for
  GFDM},'' in \emph{2014 IEEE Wireless Commun. and Networking Conf. (WCNC)},
  April 2014, pp. 1053--1058.

\bibitem{akai17}
Y.~Akai, Y.~Enjoji, Y.~Sanada, R.~Kimura, H.~Matsuda, N.~Kusashima, and
  R.~Sawai, ``{Channel estimation with scattered pilots in GFDM with multiple
  subcarrier bandwidths},'' in \emph{2017 IEEE 28th Annual International
  Symposium on Personal, Indoor, and Mobile Radio Communications (PIMRC)}, Oct
  2017, pp. 1--5.

\bibitem{ehsanfar19}
S.~{Ehsanfar}, M.~{Chafii}, and G.~{Fettweis}, ``{Time-Variant Pilot- and
  CP-Aided Channel Estimation for GFDM},'' in \emph{ICC 2019 - 2019 IEEE
  International Conference on Communications (ICC)}, 2019, pp. 1--6.

\bibitem{tai20}
C.~{Tai}, B.~{Su}, and C.~{Jia}, ``{Interference-Precancelled Pilot Design for
  LMMSE Channel Estimation of GFDM},'' in \emph{2020 IEEE 21st International
  Workshop on Signal Processing Advances in Wireless Communications (SPAWC)},
  2020, pp. 1--5.

\bibitem{ehsanfar16b}
S.~Ehsanfar, M.~Matthe, D.~Zhang, and G.~Fettweis, ``{Theoretical Analysis and
  CRLB Evaluation for Pilot-Aided Channel Estimation in GFDM},'' in \emph{2016
  IEEE Global Communications Conference (GLOBECOM)}, Dec 2016, pp. 1--7.

\bibitem{ehsanfar17}
------, ``{Interference-Free Pilots Insertion for MIMO-GFDM Channel
  Estimation},'' in \emph{2017 IEEE Wireless Communications and Networking
  Conference (WCNC)}, March 2017, pp. 1--6.

\bibitem{mirabbasi00}
S.~{Mirabbasi} and K.~{Martin}, ``{Classical and modern receiver
  architectures},'' \emph{IEEE Communications Magazine}, vol.~38, no.~11, pp.
  132--139, 2000.

\bibitem{pan12}
Y.~{Pan} and S.~{Phoong}, ``{A Time-Domain Joint Estimation Algorithm for CFO
  and I/Q Imbalance in Wideband Direct-Conversion Receivers},'' \emph{IEEE
  Transactions on Wireless Communications}, vol.~11, no.~7, pp. 2353--2361,
  2012.

\bibitem{lupupa15}
M.~Lupupa and J.~Qi, ``I/q imbalance in generalized frequency division
  multiplexing under weibull fading,'' in \emph{Personal, Indoor, and Mobile
  Radio Commun. (PIMRC), 2015 IEEE 26th Annual Int. Symposium on}, Aug 2015,
  pp. 471--476.

\bibitem{tang17b}
N.~{Tang}, S.~{He}, C.~{Xue}, Y.~{Huang}, and L.~{Yang}, ``{IQ Imbalance
  Compensation for Generalized Frequency Division Multiplexing Systems},''
  \emph{IEEE Wireless Communications Letters}, vol.~6, no.~4, pp. 422--425, Aug
  2017.

\bibitem{cheng18}
H.~{Cheng}, Y.~{Xia}, Y.~{Huang}, L.~{Yang}, and D.~P. {Mandic}, ``{A
  Normalized Complex LMS Based Blind I/Q Imbalance Compensator for GFDM
  Receivers and Its Full Second-Order Performance Analysis},'' \emph{IEEE
  Transactions on Signal Processing}, vol.~66, no.~17, pp. 4701--4712, 2018.

\bibitem{datta12a}
R.~Datta, N.~Michailow, M.~Lentmaier, and G.~Fettweis, ``{GFDM} interference
  cancellation for flexible cognitive radio {PHY} design,'' in \emph{Veh.
  Technol. Conf. (VTC Fall), 2012 IEEE}, Sep. 2012, pp. 1--5.

\bibitem{datta14b}
R.~Datta and G.~Fettweis, ``Improved aclr by cancellation carrier insertion in
  gfdm based cognitive radios,'' in \emph{2014 IEEE 79th Veh. Technol. Conf.
  (VTC Spring)}, May 2014, pp. 1--5.

\bibitem{panaitopol12}
D.~Panaitopol, R.~Datta, and G.~Fettweis, ``Cyclostationary detection of
  cognitive radio systems using gfdm modulation,'' in \emph{2012 IEEE Wireless
  Commun. and Networking Conf. (WCNC)}, April 2012, pp. 930--934.

\bibitem{danneberg14a}
M.~Danneberg, R.~Datta, and G.~Fettweis, ``Experimental testbed for dynamic
  spectrum access and sensing of 5g gfdm waveforms,'' in \emph{2014 IEEE 80th
  Veh. Technol. Conf. (VTC2014-Fall)}, Sept 2014, pp. 1--5.

\bibitem{mohammadian19}
A.~{Mohammadian}, M.~{Baghani}, and C.~{Tellambura}, ``{Optimal Power
  Allocation of GFDM Secondary Links With Power Amplifier Nonlinearity and
  ACI},'' \emph{IEEE Wireless Communications Letters}, vol.~8, no.~1, pp.
  93--96, 2019.

\bibitem{mohammadian19b}
A.~{Mohammadian} and C.~{Tellambura}, ``{Full-Duplex GFDM Radio Transceivers in
  the Presence of Phase Noise, CFO and IQ Imbalance},'' in \emph{ICC 2019 -
  2019 IEEE International Conference on Communications (ICC)}, 2019, pp. 1--6.

\bibitem{mohammadian20}
A.~{Mohammadian}, C.~{Tellambura}, and M.~{Valkama}, ``{Analysis of
  Self-Interference Cancellation Under Phase Noise, CFO, and IQ Imbalance in
  GFDM Full-Duplex Transceivers},'' \emph{IEEE Transactions on Vehicular
  Technology}, vol.~69, no.~1, pp. 700--713, 2020.

\bibitem{mohammadian19c}
A.~{Mohammadian} and C.~{Tellambura}, ``{GFDM-Modulated Full-Duplex Cognitive
  Radio Networks in the Presence of RF Impairments},'' in \emph{2019 IEEE 30th
  Annual International Symposium on Personal, Indoor and Mobile Radio
  Communications (PIMRC)}, 2019, pp. 1--6.

\end{thebibliography}

%





\end{document}